\documentclass[12pt]{article}
\usepackage{amssymb,amsmath,amsthm,latexsym}
\usepackage[utf8]{inputenc}
\usepackage{geometry}
\usepackage{yfonts}
\usepackage{xfrac}
\usepackage[c]{esvect}
\usepackage{setspace}
\usepackage{graphicx}
\usepackage{indentfirst} %indentar parágrafos
\usepackage{amsmath,amsfonts,amsthm,amscd,upref,amstext}
\usepackage{url}	
\usepackage[titletoc]{appendix}

\usepackage{array} % For the p{...} column type
\usepackage{threeparttable} % For threeparttable environment
\usepackage{caption} % For \caption
\usepackage{booktabs} % For \toprule, \midrule, \bottomrule
\usepackage[table]{xcolor} % For \textcolor
\usepackage{float} % For [H] specifier
\usepackage{amsmath} % For \textsf and other math commands
\usepackage{booktabs}
\usepackage{color,soul}
\usepackage{xcolor}
\sethlcolor{yellow}

\usepackage{tablefootnote}
\usepackage{footnote}
\usepackage{threeparttable}
\usepackage{booktabs}
\usepackage{siunitx,caption}
\usepackage{adjustbox}
\usepackage{tikz}
\usetikzlibrary{arrows.meta,automata,positioning,shadows}
\usepackage{amssymb,amsmath,amsthm,latexsym}
\usepackage{footnote}
\usepackage{graphicx}
\usepackage{float}
\usepackage{spverbatim}
\usepackage{fancybox,fancyvrb}
\newtheorem{thm}{Theorem}[section]

\newtheorem{conjec}[thm]{Mathematical Conjecture}
{\theoremstyle{remark}\newtheorem*{remark}{Remark} }

\newcommand{\keyword}[1]{\textsf{#1}}

\title{GenTwoArmsTrialSize: An R Statistical Software Package to estimate Generalized Two Arms Randomized Clinical Trial Sample Size}

\author{Mohsen Soltanifar$^{1,6}$, Chel Hee Lee$^{2,5}$, Amin Shirazi$^{3}$,\\
Martha Behnke$^{6}$, Ilfra Raymond-Loher$^{6}$, and Getachew A. Dagne$^{4}$
\footnote{$^{1}$ \quad Analytics Division, College of Professional Studies, Northeastern University, 1400-410 West Georgia Street, Vancouver, BC V6B 1Z3, Canada\\
$^{2}$ \quad  Department of Mathematics and Statistics, University of Calgary, 2500 University Drive NW, Calgary, AB T2N 1N4, Canada \\
$^{3}$ \quad Department of Statistics, College of Liberal Arts and Sciences, Iowa State University, 2438 Osborn Dr, Ames, 50011-1090, Iowa, USA\\
$^{4}$ \quad MPH Biostatistics Program, College of Public Health, University of South Florida, 3010 USF Banyan Circle, Tampa, FL, USA\\
$^{5}$ \quad Department of Critical Care Medicine, Alberta Heath Services \& University of Calgary, 3260 Hospital Dr NW, Calgary, AB T2N 4Z6, Canada\\
$^{6}$ \quad Biostatistics \& Programming Division, Biometrics Department, ClinChoice Inc.,(a) 750-2 Robert Speck Parkway, Mississauga, ON L4Z 1H8, Canada; (b) 11300 Virginia Dr., 4th Floor, Fort Washington, PA 19034, USA.}}

\date{\today}
\begin{document}
\maketitle
\doublespacing
\abstract{\noindent The precise calculation of sample sizes is a crucial aspect in the design of clinical trials particularly for pharmaceutical statisticians. While various R statistical software packages have been developed by researchers to estimate required sample sizes under different assumptions, there has been a notable absence of a standalone R statistical software package that allows researchers to comprehensively estimate sample sizes under generalized scenarios. This paper introduces the R statistical software package "GenTwoArmsTrialSize," available on the Comprehensive R Archive Network (CRAN), designed for estimating the required sample size in two-arm clinical trials. The package incorporates four endpoint types, two trial treatment designs, four types of hypothesis tests, as well as considerations for noncompliance and loss of follow-up, providing researchers with the capability to estimate sample sizes across 24 scenarios. To facilitate understanding of the estimation process and illuminate the impact of noncompliance and loss of follow-up on the size and variability of estimations, the paper includes four hypothetical examples and one applied example. The discussion encompasses the package's limitations and outlines directions for future extensions and improvements.}

\keyword{\textbf{Key Words:} R statistical software, computational biostatistics, clinical trials, sample size, compliance, follow-up}

%%%%%%%%%%%%%%%%%%%%%%%%%%%%%%%%%%%%%%%%%%%%%%%%%%

\begin{center}
\emph{With a large enough sample, any outrageous thing is likely to happen.\newline\\
\hspace{10 cm}Persi Diaconis \&  Frederick Mosteller-1989}
\end{center}

%%%%%%%%%%%%%%%%%%%%%

\section{Introduction}\label{sec1}
\unskip
This introduction is divided into five subsections as follows: Section \ref{sec1.1} provides a  brief review of the sample size and power in clinical trials. Section \ref{sec1.2} presents the statistical background required for sample size estimation in conventional two arms clinical trials. \mbox{Section \ref{sec1.3}} provides a summary of sample size calculation in R statistical software. Section \ref{sec1.4} discusses the motivations behind the ``GenTwoArmsTrialSize'' statistical software package. Finally, Section \ref{sec1.5} presents the study outline.\par 

\subsection{A Holistic View of Sample Size and Power in Clinical Trials}\label{sec1.1} 

In order for pharmaceutical companies to make informed decisions in clinical trials, it is necessary that regulators, physicians, stakeholders, and patients have a clear understanding of the risks and potential benefits of treatment(s) in clinical trials. Otherwise, lack of this clarity can result in misunderstanding of the treatment effect. To minimize lack of clarity of treatment effect, the Estimand must be constructed. An Estimand defines the target of estimation for a particular trial objective (i.e. what is to be estimated). By having a clear understanding of Estimand, proper planning can be made in how missing data should be handled. The five attributes needed to construct an Estimand are: Treatment, population, variable (endpoint), intercurrent events and how they will be handled and population-level summary  \cite{Mitroiuetall2022}. Given its significant impact on the trial design and its conduct, it has attained significant attention from the United States Food and Drug Administration, a key regulatory member of the International Council for Harmonisation of Technical Requirements for Pharmaceuticals for
Human Use (ICH) \cite{FDA2021}. In the statistical design and analysis stage, the estimand takes the form of a quantity \cite{Lundbergetall2021}. The estimand contributes core information to the calculation of the sample size without the later actually being a ‘component’ of the former.\par 

Sample size calculation in the Randomized Clinical Trials (RCT) is one of the most prominent required tasks in their designs \cite{Chowetal2018,Julious2023, Parketal2023}. There are numerous rationales for meticulous determination of sample size in clinical trial design, encompassing economic considerations (such as saving money, time, and resources), ethical considerations (such as avoiding unnecessary exposure of additional patients to futile or harmful drugs), and scientific considerations (such as guaranteeing the validity, accuracy, and reliability of the results) \cite{Parketal2023}. Two important dimensions in sample size calculations need their own special attention: (i) Relationships, (ii) Context. In the following we clarify our position in this paper in each dimension: \par 
\subsubsection{Relationships}\label{sec1.1.1}
\begin{figure}[H]
\includegraphics[clip,width=0.60\columnwidth,height=0.36\textwidth]{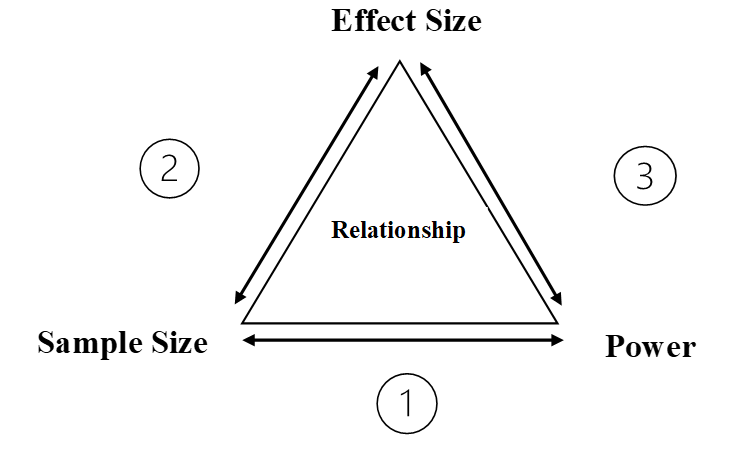}%
\caption{ Three way relationships between the key elements of sample size calculation\label{fig1}}
\end{figure}

First, in the broader field of calculating sample size, aside from significance level, three key components come into play: (i) the sample size itself, (ii) the effect size, and (iii) the power. These give rise to three interrelated relationships outlined as follows \cite{Chowetal2018} (Figure \ref{fig1}):\par
\begin{enumerate}
    \item Sample Size in terms of Power and Effect Size(Re 1 \& Re 2),
    \item Power in terms of Effect size and Sample Size(Re 1 \& Re 3),
    \item Effect Size in terms of Power and Sample Size (Re 2 \& Re 3).
\end{enumerate}

\vspace{-6pt}

\subsubsection{Context}\label{sec1.1.2}
Second, in the same broader field of sample size calculation,  there are several contexts to consider including but not limited to \cite{Chowetal2018} :\par 
\begin{enumerate}
    \item Adaptive vs. Fixed,
    \item Parallel vs. Crossover,
    \item Longitudinal vs. Cross-sectional,
    \item Equality/Noninferiority/Superiority/Equivalence,
    \item One Arm/Two Arms/Multiple Arms,
    \item Single structure vs. Cluster,
    \item Continuous/Binary/Time to Event/Ordinal Categorical end points
    \item Frequentist vs. Bayesian
    \item Parametric vs. Non parametric,
    \item One Stage/Two Stages/Multiple Stages,
    \item Single vs. Multilevel.
\end{enumerate}
In this paper, we consider sample size calculation in the first relationship within the following context: (i) Fixed, (ii) Cross-sectional, (iii) Two Arms, (iv) Single structure, (v) Frequentist, (vi) Parametric, (vii) One stage and (viii) Single level. \par  
\subsection{Statistical Background}\label{sec1.2}  
In this section we first review briefly the mathematical statistics background for sample size calculation of conventional two arms trials by their end point from Intention To Treat (ITT) perspective. Then, we consider the issues of noncompliance and expected loss of follow up. We conclude with the secondary ITT sample size calculations in terms of noncompliance information and the Complier Average Causal Effect (CACE) information.\par   
\subsubsection{Sample Size Estimations by Trial End Points: Primary ITT}\label{sec1.2.1}
We consider four main trial end points: Continuous, Binary(large samples), Time to Event (TTE) and Ordinal Categorical  and briefly present the statistical background for each case given underlying assumptions for involved populations as in \cite{Chowetal2018, Shih2022,Campbelletal1995}:\par 
Let $\theta$ be the parameter of interest denoting either $\mu,$ $p,$ $-\lambda,$ and $\psi,$ as the mean of continuous outcome, proportion of the binary outcome, negative hazard of exponential model of time to event outcome and logarithm of odds of cumulative probability of the  ordinal categorical outcome, respectively. Furthermore, let  $\theta_1$ and $\theta_2$ denote the parameter of interest for the control arm and the treatment arm of the clinical trial, respectively. Then, for given margin $\delta>0$ there are four conventional two-sided statistical tests of interest at significant level $\alpha$ (two-sided test at an $\alpha$ level, or a one-sided test at an $\alpha/2$ level) and power $1-\beta$ as follows: \par 
\begin{eqnarray}
Equality &:&  H_0: \theta_2-\theta_1=0\ \ \ \ \ vs.\ \ \     H_1: \theta_2-\theta_1\neq0,\label{eq1} \\
Noninferiority &:&  H_0: \theta_2-\theta_1\leq -\delta\ \ \ vs.\ \ \     H_1: \theta_2-\theta_1> -\delta,\label{eq2}  \\
Superiority &:&  H_0: \theta_2-\theta_1\leq +\delta\ \ \ vs.\ \ \     H_1: \theta_2-\theta_1> +\delta,\label{eq3}  \\
Equivalence &:&  H_0: |\theta_2-\theta_1|\geq +\delta\ \ \ vs.\ \ \     H_1: |\theta_2-\theta_1|<+\delta.\label{eq4}
\end{eqnarray}
Sample size estimations for the hypothesis tests (\ref{eq1})-(\ref{eq4}) are as follows \cite{Chowetal2018, Shih2022,Whitehead1993}:\newline  
\indent First of all, we consider the two arms parallel or crossover clinical trial with continuous outcome where in which $n_{2}$ and $n_{1}$ are the required number of patients for the treatment arm and the control arm, respectively. Given assumption $n_{1}=kn_{2} (k>0),$ the required number of patients is \cite{Chowetal2018}:\par 
\begin{eqnarray}\label{eq5}
n_{2}(\mu_1,\mu_2)&=& \inf \{ m\in \mathbb{N}| 1-T_{(1+k)m-2}(t_{U(\alpha),(1+k)m-2}|\sqrt{\frac{sm}{\sigma^2(1+k^{-1})}}V(\mu_1,\mu_2,\delta))=W(\beta)\}\nonumber\\
&&
\end{eqnarray}
where in which $T_n(.|M)$ is the Cumulative Distribution Function (CDF) of the non-central t-student distribution with $n$ degrees of freedom with non-centrality parameter of $M.$ Also, for special case of large sample size (e.g., $n>1000$) or known population variance (e.g., $\sigma^2=10$ ) we have $T_{(1+k)m-2}(t_{U(\alpha),(1+k)m-2}|\sqrt{\frac{sm}{\sigma^2(1+k^{-1})}}V(\mu_1,\mu_2,\delta))\approx \Phi_Z(z_{U(\alpha)}-\sqrt{\frac{sm}{\sigma^2(1+k^{-1})}}V(\mu_1,\mu_2,\delta)),$ yielding to explicit closed form formulae for $n_{2}$.  Here, for the case of parallel and crossover treatment allocations we have $(k,s):=(k,1)$ and $(k,s):=(1,4),$ respectively. Finally, given the hypothesis testing of equality, noninferiority, superiority, and equivalence, we have $U(\alpha)=\alpha/2,\alpha,\alpha,\alpha;$ $V(\mu_1,\mu_2,\delta)=|\mu_2-\mu_1|, \mu_2-\mu_1-\delta, \mu_2-\mu_1-\delta, \delta-|\mu_2-\mu_1|;$ and $W(\beta)=1-\beta,1-\beta,1-\beta, 1-\beta/2,$ respectively. \par 
Second, we consider the two arms parallel or crossover clinical trial for large samples with binary outcome. With similar assumptions on the arms related required number of patients as in the continuous outcome, the required number of patients is \cite{Chowetal2018}:\par 
\begin{eqnarray}\label{eq6}
n_{2}(p_1,p_2)&=& \inf \{ m\in \mathbb{N}| 1-\Phi_Z(z_{U(\alpha)}-\sqrt{\frac{sm}{\sigma_{d,p_1,p_2,k}^2}}V(p_1,p_2,\delta))=W(\beta)\}\nonumber\\
&&
\end{eqnarray}
where  for the case of parallel and crossover treatment allocations we have $(\sigma_{d,p_1,p_2,k}^2,s):=(p_1(1-p_1)k^{-1}+p_2(1-p_2),1)$ and $(\sigma_{d,p_1,p_2,k}^2,s):=(\sigma_d^2,2),$ in which $\sigma_d^2$ refers to variance of difference of two arms, respectively. Here, similar to the continuous outcome,  given the hypothesis testing of equality, noninferiority, superiority, and equivalence, we have $U(\alpha)=\alpha/2,\alpha,\alpha,\alpha;$ $V(p_1,p_2,\delta)=|p_2-p_1|, p_2-p_1-\delta, p_2-p_1-\delta, \delta-|p_2-p_1|;$ and $W(\beta)=1-\beta,1-\beta,1-\beta, 1-\beta/2,$ respectively. \par
\begin{remark}\label{smallsamplesize}
The sample size calculation for parallel clinical trials with small sample size (e.g., $n_i<((\sqrt{(1-p_i)/p_i}-\sqrt{p_i/(1-p_i)})/0.3)^2: \ \ i=1,2$ \cite{Boxetal2005}) involves continuity correction. Here, it is sufficient to replace the term $V(p_1,p_2,\delta)$ in equation (\ref{eq6}) with $V(p_1,p_2,\delta,m,k)$ defined by $V(p_1,p_2,\delta,m,k)=V(p_1,p_2,\delta)-(1+1/k)/(2m)$ \cite{Julious2023}. An approximate solution for equation (\ref{eq6}) by setting $A=(1+1/k)/2, B=\sigma_{d,p_1,p_2,k}, C=z_{U(\alpha)}+z_{(1-W(\beta))},$ and $ D=V(p_1,p_2,\delta)$ is $n_{2}(p_1,p_2)\approx ((BC+\sqrt{(BC)^2+4AD})/(2D))^2.$
\end{remark}
Third, we consider the two arms parallel clinical trial  with time to event outcome. With similar assumptions on the arms related required number of patients as in the continuous outcome, the required number of patients is \cite{Chowetal2018}:\par 
\begin{eqnarray}\label{eq7}
n_{2}(\lambda_1,\lambda_2)&=& \inf \{ m\in \mathbb{N}| 1-\Phi_Z(z_{U(\alpha)}-\sqrt{\frac{m}{\sigma^2(\lambda_1)k^{-1}+\sigma^2(\lambda_2)}}V(\lambda_1,\lambda_2,\delta))=W(\beta)\}\nonumber\\
&&
\end{eqnarray}
where, $\sigma^2(\lambda_i)\ \ (i=1,2)$ are the associated variances in the Central Limit Theorem asymptotic estimations of $\lambda_i\ \ (i=1,2),$ respectively. Similar to the continuous outcome,  given the hypothesis testing of equality, noninferiority, superiority, and equivalence, we have $U(\alpha)=\alpha/2,\alpha,\alpha,\alpha;$ $V(\lambda_1,\lambda_2,\delta)=|\lambda_1-\lambda_2|, \lambda_1-\lambda_2-\delta, \lambda_1-\lambda_2-\delta, \delta-|\lambda_1-\lambda_2|;$ and $W(\beta)=1-\beta,1-\beta,1-\beta, 1-\beta/2,$ respectively. \par
Finally, we consider the two arms parallel clinical trial  with ordinal categorical outcome. With similar assumptions on the arms related required number of patients as in the continuous outcome, the required number of patients is \cite{ Shih2022,Whitehead1993,Julious2023}:\par 
\begin{eqnarray}\label{eq8}
n_{2}(\psi_1,\psi_2)&=& \inf \{ m\in \mathbb{N}| 1-\Phi_Z(z_{U(\alpha)}-\sqrt{\frac{mk}{3(k+1)}(1-\sum_{i=1}^{J}\overline{p}_i^3)}V(\psi_1,\psi_2,\delta))=W(\beta)\}\nonumber\\
&&\overline{p}_j=(p_{1j}+p_{2j})/2: (1\leq j\leq J)
\end{eqnarray}
where $p_{ij}\ \ (i=1,2, j=1,\cdots,J),$ is the probability of outcome in the $j$th category of the $i$th trial arm. Similar to the continuous outcome,  given the hypothesis testing of equality, noninferiority, superiority, and equivalence, we have $U(\alpha)=\alpha/2,\alpha,\alpha,\alpha;$ $V(\psi_1,\psi_2,\delta)=|\psi_2-\psi_1|, (\psi_2-\psi_1)-\delta, (\psi_2-\psi_1)-\delta, \delta-|\psi_2-\psi_1|;$ and $W(\beta)=1-\beta,1-\beta,1-\beta, 1-\beta/2,$ respectively (Note that $\log(OR)=\psi_2-\psi_1$). \par

\subsubsection{Noncompliance and Expected Loss of Follow-up}\label{sec1.2.2}
 
Sample size estimation based on ITT are considered the ideal estimations. However, in the reality in most clinical trials the perfect design of the trial is violated during the trial run. Two prominent cases of violation of trial original design are noncompliance and loss of follow-up described as follows \cite{Wittes2002,Jo2002, Sato2000, Lachin1986, Shih2022}: \par 
Noncompliance refers to the scenario that trial patients fail to adhere to their original assigned trial treatment arms regimen in the study protocol (e.g., missing a fraction of assigned medication/drug for the entire trial). Let $\theta$ be a parameter of interest (i.e., $\theta=\mu,p,-\lambda,\log(\psi)$) and consider the $\theta_1,\theta_2$ the ITT mean estimations of the trial control arms and treatment arms outcomes, respectively. Assume, $\rho_1,\rho_2$ to be drop-out proportions in the trial control arm  and treatment arm, respectively. Under conditions where treatments are switched (e.g., as occurs in any open label active control trial with market available active control agent, or in some oncology trials) \cite{Shih2022,Jo2002}, and under the noncompliance assumption the expected mean endpoint estimation of the trial control arm and treatment arm equals to \cite{Sato2000,Lachin1986,Shih2022}:
\begin{eqnarray}
\theta_1^{*}&=&(1-\rho_1)\theta_1+\rho_1\theta_2,\label{eq9}\\
\theta_2^{*}&=&\rho_2\theta_1+(1-\rho_2)\theta_2.\label{eq10}
\end{eqnarray}
Consequently, the secondary  ITT  estimation in terms of the probability of compliance and the Complier Average Causal Effect (CACE) is given by (See Appendix \ref{secA}):
\begin{eqnarray}\label{eq11}
(\theta_2^{*}-\theta_1^{*})_{ITT}&=& (1-\rho_2-\rho_1)_{Prob(Compliance)}\times(\theta_2-\theta_1)_{CACE}.   
\end{eqnarray}
In particular, for the ideal scenario of the perfect compliance $\rho_1=\rho_2=0,$ the secondary ITT estimations in equations (\ref{eq9})-(\ref{eq11}) match to the primary ITT estimations.\par 
Loss of follow-up refers to the scenario in which the patients do not complete the trial and their trial outcomes are untraceable. Let $r$ be the projected pooled proportion of patients in both arms that are lost to follow-up by the end of trial and $n$ be the unadjusted required number of patients for completion of trial. Then, under both loss of follow-up and no imputation of the missing values in the trial assumptions, the adjusted required number of patients for the completion of the trial equals to \cite{Shih2022}:
\begin{eqnarray}\label{eq12}
n^*&=&\frac{n}{1-r}.    
\end{eqnarray}

\subsubsection{Sample Size Estimations by Trial End Points: Secondary ITT}\label{sec1.2.3}
 
In many clinical trials, both noncompliance and loss of follow-up occur yielding to adjustments for both scenarios. Hence, considering equations (\ref{eq9},\ref{eq10},\ref{eq12}) in the ITT equations the secondary ITT estimation is given by:
$$n^{*}(\theta_1,\theta_2,\rho_1,\rho_2,r)=\frac{n(((1-\rho_1)\theta_1+\rho_1\theta_2),(\rho_2\theta_1+(1-\rho_2)\theta_2))}{1-r}.$$
In particular, considering equations (\ref{eq9},\ref{eq10},\ref{eq12}) in the ITT equations (\ref{eq5},\ref{eq6},\ref{eq7},\ref{eq8}) the secondary ITT estimations are given by: 
\begin{eqnarray}
n_{2}^{*}(\mu_1,\mu_2,\rho_1,\rho_2,r)&=&\frac{n_{2}(((1-\rho_1)\mu_1+\rho_1\mu_2),(\rho_2\mu_1+(1-\rho_2)\mu_2))}{1-r},\label{eq13}\\
n_{2}^{*}(p_1,p_2,\rho_1,\rho_2,r)&=&\frac{n_{2}(((1-\rho_1)p_1+\rho_1 p_2),(\rho_2p_1+(1-\rho_2)p_2))}{1-r},\label{eq14}\\
n_{2}^{*}(\lambda_1,\lambda_2,\rho_1,\rho_2,r)&=&\frac{n_{2}(((1-\rho_1)\lambda_1+\rho_1\lambda_2),(\rho_2\lambda_1+(1-\rho_2)\lambda_2))}{1-r},\label{eq15}\\
n_{2}^{*}(\psi_1,\psi_2,\rho_1,\rho_2,r)&=&\frac{n_{2}(((1-\rho_1)\psi_1+\rho_1 \psi_2),(\rho_2\psi_1+(1-\rho_2)\psi_2))}{1-r},\label{eq16}\\
&&\overline{p}_i^{*}=((1-\rho_1+\rho_2)p_{1i}+(1+\rho_1-\rho_2)p_{2i})/2: (1\leq i\leq J).\nonumber
\end{eqnarray}

\subsection{Sample Size Calculation in Noncommercial statistical software}\label{sec1.3} 

Several R statistical packages have been presented for sample size calculation for parametric two arms clinical trials \cite{Chowetal2018,Parketal2023,Chenetal2017, Zhangetall2020, Harrell2023}. The most prominent and comprehensive available package is "TrialSize" \cite{Zhangetall2020} accompanied with 15 functions. 
Other additional estimation R packages and functions -presented as R-package name(Function)- are as follows \cite{Harrell2023, Parketal2023,Chenetal2017,Scherer2022,Robinetall2011,Iddi2022, Donohueetall2021,Liu1997,Lu2008,Hu2021}: 
(i) continuous endpoint: Stats(power.t.test), samplesize(n.indept.t.test.eq), samplesize(n.indept.t.test.neq), samplesize(n.paired.t.test), samplesize(n.welch.test), pwr(pwr.t.test), Longpower(liu.liang.linear.power), longpower(lmmpower); (ii) binary endpoint: gsDesign(nBinomial), pwr(pwr.p.test), Stats(power.prop.test),  pwr(pwr.2p.test); (iii) time to event endpoint:  gsDesign(nSurvival);
and, (iv) ordinal endpoint: Hmisc(posamsize), samplesize(n.wilcox.ord).\par 
Apart from R, there exist other notable noncommercial statistical software options for calculating sample sizes, such as STPLAN, which utilizes code derived from Fortran 77 \cite{Brownetal2010}.\par 

\subsection{Motivation}\label{sec1.4} 
Until late 2023 several statistical packages including those mentioned above were proposed by the researchers to estimate required sample size for conventional two arms trials. However, the authors believe that more improvement is possible given the following reasons:\par 
First, no developed R package till now has considered noncompliance and loss of follow up in the context of frequentist estimations. The only available one is limited to context of simulation based adaptive bayesian setting \cite{Dmitrienko2023}. This lack of consideration in the design of clinical trials is crucial given the prominent role of  compliance and perfect patient follow up in ensuring safety and efficacy of the involved treatment in the new trial \cite{Besch1995,Czobor2011,Cuzick1997,Boudes1998}. \par
Second, to authors best knowledge, there is no R package yet to estimate required sample size for two arms parallel -ordinal categorical endpoint designs with noninferioirty, superiority and equivalence hypothesis tests. The only available R package Hmisc for two arms parallel ordinal categorical endpoint design merely considers the equality hypothesis test \cite{Harrell2023}.\par 
Third, to estimate sample size for different scenarios of the conventional two arms trial designs one need to use different R packages with different input functions and parameters as shown by those mentioned in above \cite{Chowetal2018,Parketal2023,Scherer2022, Chenetal2017,Harrell2023}. This situation makes estimation process sub optimal given heterogeneity in the packages update dates, their available functions, and their  available parameters.\par 
Finally, with STPLAN, certain constraints are evident, including its omission of arm noncompliance rates and loss of follow-up rates in the most scenario calculations (except count outcomes), its omission of equivalence scenarios in hypothesis testing, and its absence of support for crossover designs in calculations \cite{Brownetal2010}. \par 
In this paper, we present R package ``GenTwoArmsTrialSize" \cite{SoltanifarLee2023} that simultaneously addresses all of above limitations and issues. \par 

\subsection{Study Outline}\label{sec1.5} 

This paper is organized as follows. We start by providing minimal basic instructions for installing the GenTwoArmsTrialSize package. The main four functions for sample size calculations for continuous, binary, time to event and ordinal categorical endpoints are then illustrated. Then, we present four worked examples study (one per end point type) for  estimation of required sample size given noncompliance and loss of follow-up in each setting using the package. Next, variation of sample size estimation by noncompliance rates and loss of follow up rates are discussed by presenting an application in trial design and then, the role of software is explored. Finally, we discuss the contributions of the GenTwoArmsTrialSize package to the designs of clinical trials, its current limitations, and prospective for its future developments.\par

\section{The GenTwoArmsTrialSize Package}\label{sec2}

\subsection{The Background and Installation}\label{sec2.1}
The ``GenTwoArmsTrialSize'' package is based on the two current well-known R packages for sample size calculations in clinical trials ``TrialSize" \cite{Zhangetall2020} and ``Hmisc" \cite{Harrell2023} and the R package for the general grammar of data manipulation ``dplyr" \cite{Wickhametall2023}. Its estimation process has the following features: \par  
\begin{enumerate}
    \item It generalizes all 20 two arm scenarios in the package ``TrialSize" in terms of additional parameters: Noncompliance and Loss of follow up.
    \item It adds three more scenarios to the only available single scenario in package ``Hmisc" for the two arm Ordinal case with the total of 4 scenarios.
    \item It generalizes all combined 4 two arm scenarios in the previous item in terms of additional parameters: Noncompliance and Loss of follow up.
    \item It combines 6 functions in each of scenarios 1--8,  6 functions in scenarios 9--16, and 3 functions in scenarios 17--20 to merely one function in each range of scenarios, respectively. 
\end{enumerate}

The estimation process presented above may involve  4 $\times$ 2 $\times$ $4=32$ potential scenarios. Table \ref{table1} presents 24  scenarios offered by the package GenTwoArmsTrialSize (version 0.0.5). \par 
The package (version 0.0.5)  is available from the Comprehensive R Archive Network (CRAN) at  \url{https://CRAN.R-project.org/package=GenTwoArmsTrialSize} (accessed on 11 December 2023). The following code chunk shows how to install and load the package:
\par    
\begin{Verbatim}
line #1:> install.packages("GenTwoArmsTrialSize", dependencies=TRUE)
line #2:> library(GenTwoArmsTrialSize)
\end{Verbatim}
   
The ``GenTwoArmsTrialSize'' package involves four major multi-variate functions \textsf{getSizeMean()},  \textsf{getSizeProp()}, \textsf{getSizeTTE} and \textsf{getSizeOrd()}. Details are described in the coming sections. The reader may try to type \textsf{?getSizeMean()},  \textsf{?getSizeProp()}, \textsf{?getSizeTTE} or \textsf{?getSizeOrd()}  in R for the details.\par

\begin{table}[H]
\caption{A summary of potential estimation scenarios by R package ``GenTwoArmsTrialSize'' (version 0.0.5)).\label{table1}} %title of the table
\scriptsize{
\begin{threeparttable}[b]
\centering % centering table
\begin{adjustbox}{max width=\textwidth,
max height=\textheight}
\begin{tabular}{p{1.3cm}p{1.7cm}p{1.5cm} p{3.5cm}p{1.5cm}p{1.5cm}p{2.5cm}} % creating four columns
\toprule % inserts single-line
\textbf{Scenario} & \textbf{Endpoint} & \textbf{Design} & \textbf{Hypothesis. Test}   & \textbf{Noncom\Large{${*}$}} & \textbf{LoF \Large{${*}$}} & \textbf{Function\large{${\dagger}$}}\\ % Entering row contents
\midrule % inserts single-line
1 & Continuous &Parallel & Equality &  \textcolor{blue}{$Yes^{*}$} & \textcolor{blue}{$Yes^{*}$} & \footnotesize{\textsf{getSizeMean()}}\\
2 &            &         & Noninferiority & \textcolor{blue}{$Yes^{*}$} & \textcolor{blue}{$Yes^{*}$} & \footnotesize{\textsf{getSizeMean()}}\\
3 &            &         & Superiority& \textcolor{blue}{$Yes^{*}$}  & \textcolor{blue}{$Yes^{*}$} & \footnotesize{\textsf{getSizeMean()}}\\
4 &            &         & Equivalence & \textcolor{blue}{$Yes^{*}$} & \textcolor{blue}{$Yes^{*}$} & \footnotesize{\textsf{getSizeMean()}}\\
5 &  &Crossover & Equality &  \textcolor{blue}{$Yes^{*}$} & \textcolor{blue}{$Yes^{*}$} & \footnotesize{\textsf{getSizeMean()}}\\
6 &  & & Noninferiority & \textcolor{blue}{$Yes^{*}$} & \textcolor{blue}{$Yes^{*}$} & \footnotesize{\textsf{getSizeMean()}}\\
7 &  & & Superiority& \textcolor{blue}{$Yes^{*}$}  & \textcolor{blue}{$Yes^{*}$} & \footnotesize{\textsf{getSizeMean()}}\\
8 &  &  & Equivalence & \textcolor{blue}{$Yes^{*}$} & \textcolor{blue}{$Yes^{*}$} & \footnotesize{\textsf{getSizeMean()}}\\
9 & Binary &Parallel & Equality &  \textcolor{blue}{$Yes^{*}$} & \textcolor{blue}{$Yes^{*}$} & \footnotesize{\textsf{getSizeProp()}}\\
10 &            &         & Noninferiority & \textcolor{blue}{$Yes^{*}$} & \textcolor{blue}{$Yes^{*}$} & \footnotesize{\textsf{getSizeProp()}}\\
11 &            &         & Superiority& \textcolor{blue}{$Yes^{*}$}  & \textcolor{blue}{$Yes^{*}$} & \footnotesize{\textsf{getSizeProp()}}\\
12 &            &         & Equivalence & \textcolor{blue}{$Yes^{*}$} & \textcolor{blue}{$Yes^{*}$} & \footnotesize{\textsf{getSizeProp()}}\\
13 &  &Crossover & Equality &  \textcolor{blue}{$Yes^{*}$} & \textcolor{blue}{$Yes^{*}$} & \footnotesize{\textsf{getSizeProp()}}\\
14 &  & & Noninferiority & \textcolor{blue}{$Yes^{*}$} & \textcolor{blue}{$Yes^{*}$} & \footnotesize{\textsf{getSizeProp()}}\\
15 &  & & Superiority& \textcolor{blue}{$Yes^{*}$}  & \textcolor{blue}{$Yes^{*}$} & \footnotesize{\textsf{getSizeProp()}}\\
16 &  &  & Equivalence & \textcolor{blue}{$Yes^{*}$} & \textcolor{blue}{$Yes^{*}$} & \footnotesize{\textsf{getSizeProp()}}\\
17 & TTE &Parallel & Equality &  \textcolor{blue}{$Yes^{*}$} & \textcolor{blue}{$Yes^{*}$} & \footnotesize{\textsf{getSizeTTE()}}\\
18 &            &         & Noninferiority & \textcolor{blue}{$Yes^{*}$} & \textcolor{blue}{$Yes^{*}$} & \footnotesize{\textsf{getSizeTTE()}}\\
19 &            &         & Superiority& \textcolor{blue}{$Yes^{*}$}  & \textcolor{blue}{$Yes^{*}$} & \footnotesize{\textsf{getSizeTTE()}}\\
20 &            &         & Equivalence & \textcolor{blue}{$Yes^{*}$} & \textcolor{blue}{$Yes^{*}$} & \footnotesize{\textsf{getSizeTTE()}}\\
21 &  &Crossover & Equality &  -- & -- & \footnotesize{\textsf{--}}\\
22 &  &          & Noninferiority & -- & -- & \footnotesize{\textsf{--}}\\
23 &  &          & Superiority & --  & -- & \footnotesize{\textsf{--}}\\
24 &  &          & Equivalence & -- & -- & \footnotesize{\textsf{--}}\\
25 & Ordinal &Parallel & Equality &  \textcolor{blue}{$Yes^{*}$} & \textcolor{blue}{$Yes^{*}$} & \footnotesize{\textsf{getSizeOrd()}}\\
26 &            &         & Noninferiority & \textcolor{blue}{$Yes^{*}$} & \textcolor{blue}{$Yes^{*}$} & \footnotesize{\textcolor{blue}{\textsf{getSizeOrd()$^{\dagger}$}}}\\
27 &            &         & Superiority& \textcolor{blue}{$Yes^{*}$} & \textcolor{blue}{$Yes^{*}$} & \footnotesize{\textcolor{blue}{\textsf{getSizeOrd()$^{\dagger}$}}}\\
28 &            &         & Equivalence & \textcolor{blue}{$Yes^{*}$} & \textcolor{blue}{$Yes^{*}$} & \footnotesize{\textcolor{blue}{\textsf{getSizeOrd()$^{\dagger}$}}}\\
29 &  &Crossover & Equality &  -- & -- & \footnotesize{\textsf{--}}\\
30 &  &          & Noninferiority & -- & -- & \footnotesize{\textsf{--}}\\
31 &  &          & Superiority & --  & -- & \footnotesize{\textsf{--}}\\
32 &  &          & Equivalence & -- & -- & \footnotesize{\textsf{--}}\\
\bottomrule % inserts single-line
\end{tabular}
\end{adjustbox}
\end{threeparttable}
\begin{tablenotes}
\scriptsize
\item[a] Note: Noncom: Noncompliance; LoF: Loss of follow-up; $-$: Scenario under development; $^{*}$: New added feature; $^{\dagger}$ New added function.
\end{tablenotes}
}
\end{table}

\subsection{Sample size Calculation for Two Arms Clinical Trials with Continuous Endpoints}\label{sec2.2}

The function \textsf{getSizeMean()} calculates the required sample size for two arms clinical trial with continuous end points and needs the following ten input parameters to estimate the plausible sample size:\par 
For a given two arms trial, \textsf{design} is the allocation method (parallel/crossover), \textsf{test} is the hypothesis test(equality/noninferiority/superiority/equivalence), \textsf{alpha} is the level of test significance, \textsf{beta} is the type II error (1-power), \textsf{sigma} is the pooled standard deviation of two groups, \textsf{k} is the ratio of control arm size to the treatment arm size, \textsf{delta} is the delta margin in the test hypothesis, \textsf{TTE} is the target treatment effect, \textsf{rho} is a vector of length two with noncompliance rates of control arm and treatment arm, and,  \textsf{r} is the projected proportion of trial uniform loss of follow-up.  The function general code in R is: \par 
\begin{center}
\begin{Verbatim} 
getSizeMean(design,test,alpha, beta, sigma, k, delta, TTE, rho,r)
\end{Verbatim}    
\end{center}
As the first working example, we consider the Example 3.2.4 \cite{Chowetal2018} with additional features of noncompliance and expected loss of follow up:\par 
\textbf{Example 1.} A pharmaceutical company is contemplating a clinical trial to assess the therapeutic equivalence of two cholesterol-lowering medications in treating patients with Chronic Heart Disease (CHD). The trial will employ a parallel design, with a primary focus on measuring low-density lipoprotein (LDL) levels. The principal investigator (PI) is currently in the process of determining the sample size needed to achieve 80\% power at a significance level of 5\%. This calculation assumes a standard deviation (SD) of 10\%, a true mean difference of 0.01, and an equivalence limit of 0.05. The PI is factoring in equal-sized arms, considering noncompliance rates of 5\% for the control arm and 7\% for the treatment arm. Additionally, an overall trial loss to follow-up of 10\% is being taken into account by the PI. The R code for sample size calculations given above trial information is given by:
\begin{Verbatim} 
line #1:>  getSizeMean(design="parallel",
                        test="equivalence",
                        alpha=0.05, 
                        beta=0.20,
                        sigma=0.10,
                        k=1,
                        delta=0.05,
                        TTE=0.01, 
                        rho=c(0.05, 0.07),
                        r=0.1)
\end{Verbatim} 
yielding to following output estimations:
\begin{Verbatim}
line #1:>       n_2 n_1
line #2:> Size   113   113    
\end{Verbatim}
Hence, the PI needs to consider two arms each of size 113. We note that this sample size is larger than 108 for the case with perfect compliance and no loss of follow up. \par  

\subsection{Sample size Calculation for Two Arms Clinical Trials with Binary Endpoints}\label{sec2.3}

The function \textsf{getSizeProp()} calculates the required sample size for two arms clinical trial with binary end points with large samples and needs the following eleven input parameters to estimate the plausible sample size:\par 
For a given two arms trial, \textsf{design} is the allocation method (parallel/crossover), \textsf{test} is the hypothesis test(equality/noninferiority/superiority/equivalence), \textsf{alpha} is the level of test significance, \textsf{beta} is the type II error (1-power), \textsf{varsigma} is a vector of size two which is probability of mean response in
control and treatment arms $(varsigma1 > 0, varsigma2 > 0) := (p1, p2)$; or pooled standard deviation of two treatment arms or their difference $(varsigma1 > 0, varsigma2 > 0) := (sigma, sigma): (sigma>0), $ \textsf{k} is the ratio of control arm size to the treatment arm size, \textsf{seqnumber} Number of crossover sequences(0 if parallel;  $1+$ if crossover), \textsf{delta} is the delta margin in the test hypothesis, \textsf{TTE} is the target treatment effect, \textsf{rho} is a vector of length two with noncompliance rates of control arm and treatment arm, and,  \textsf{r} is the projected proportion of trial uniform loss of follow-up (Note that the value $1+$ refers to any value chosen from the sequence $1(AB,BA), 2(ABAB,BABA),3,\cdots$.). The function general code in R is: \par 
\begin{center}
\begin{Verbatim} 
getSizeProp(design,test,alpha,beta,varsigma,k,seqnumber,delta,TTE,rho,r)
\end{Verbatim}    
\end{center}
As the second working example, we consider the Example 4.3.4 \cite{Chowetal2018} with additional features of noncompliance and expected loss of follow up:\par
\textbf{Example 2.} A pharmaceutical company is contemplating an open label randomized crossover clinical trial to compare an inhaled insulin formulation manufactured for commercial usage for patients with type I diabetes (A) to the inhaled insulin formulation utilized in phase III clinical trials (B). The replicated crossover design consisting of two sequences of ABAB and BABA is recommended with a primary focus on comparing the safety profiles between the two formulations in terms of the incidence rate of adverse events. The principal investigator (PI) is currently in the process of determining the sample size needed to achieve 80\% power at a significance level of 5\%. This calculation assumes a difference of samples standard deviation (SD) of 50\%, no difference in incidence rates of adverse events, and a superiority limit of 10\%. The PI is factoring in equal-sized arms, considering noncompliance rates of 5\% for the control arm and 7\% for the treatment arm. Additionally, an overall trial loss to follow-up of 10\% is being taken into account by the PI. The R code for sample size calculations given above trial information is given by:
\begin{Verbatim} 
line #1:>  getSizeProp(design="crossover",
                        test="superiority",
                        alpha=0.05, 
                        beta=0.20,
                        varsigma=c(0.50,0.50),
                        k=1,
                        seqnumber=2,
                        delta=0.10,
                        TTE=0.0, 
                        rho=c(0.05, 0.07),
                        r=0.1)
\end{Verbatim} 
yielding to following output estimations:
\begin{Verbatim}
line #1:>       n_2 n_1
line #2:> Size   86   86    
\end{Verbatim}
Hence, the PI needs to consider two arms each of size 86. We note that this sample size is larger than 78 for the case with perfect compliance and no loss of follow up. \par

\subsection{Sample size Calculation for Two Arms Clinical Trials with Time-to-Event Endpoints}\label{sec2.4}

The function getSizeTTE() calculates the required sample size for two arms clinical trial with time to event end points and needs the following twelve input parameters to estimate the plausible sample size:\par   
For a given two arms trial, \textsf{design} is the allocation method (parallel/crossover), \textsf{test} is the hypothesis test(equality/noninferiority/superiority/equivalence), \textsf{alpha} is the level of test significance, beta is the type II error (1-power), \textsf{varlambda} a vector of size two $(varlambda1>0,varlambda2>0):=(lam1,lam2)$ denotes the hazard rates in control and treatment arms, \textsf{k} is the ratio of control arm size to the treatment arm size, \textsf{ttotal} is the total trial time (ttoal>0), \textsf{taccrual} is the accrual time period (taccrual>0), \textsf{gamma} is the parameter of exponential distribution (gamma>=0), \textsf{delta} is the delta margin in the test hypothesis, \textsf{rho} is a vector of length two with noncompliance rates of control arm and treatment arm, and, \textsf{r} is the projected proportion of trial uniform loss of follow-up. The function general code in R is:\par 
\begin{center}
\begin{Verbatim} 
getSizeTTE(design,test,alpha,beta,varlambda,k,ttotal,taccrual,gamma,delta,rho,r)
\end{Verbatim}    
\end{center}
As the third working example, we consider the Example 7.2.4 \cite{Chowetal2018} with additional features of noncompliance and expected loss of follow up:\par
\textbf{Example 3.} A pharmaceutical company is contemplating a randomized parallel clinical trial to compare equality of survivorship among the trial cancer survival patients in terms of time to leukemia. The two trial arms include those patients with allogeneic  transplant from an HLA-matched sibling donor (allo)  and those with an autologous transplant where their own marrow has been cleansed and returned to them after high dose of chemotherapy (auto).  The principal investigator (PI) is currently in the process of determining the sample size needed to achieve 80\% power at a significance level of 5\%. The trial is planned to last for three  years with one year accrual. Uniform patient entry for both allo and auto transplant groups is assumed. It is also assumed that the leukemia-free hazard rates for allo and auto transplant are given by 1 and 2, respectively.  The PI is factoring in equal-sized arms, considering noncompliance rates of 5\% for the control arm and 7\% for the treatment arm. Additionally, an overall trial loss to follow-up of 10\% is being taken into account by the PI. The R code for sample size calculations given above trial information is given by:\par 
\begin{Verbatim} 
line #1:>  getSizeTTE(design="parallel",
                      test="equality",
                      alpha=0.05,
                      beta=0.20,
                      varlambda=c(1,2),
                      k=1,
                      ttotal=3,
                      taccrual=1, 
                      gamma=0.00001,
                      delta=0,
                      rho=c(0.05, 0.07),
                      r=0.1)
\end{Verbatim} 
yielding to following output estimations:
\begin{Verbatim}
line #1:>       n_2 n_1
line #2:> Size   56   56    
\end{Verbatim}
Hence, the PI needs to consider two arms each of size 56. We note that this sample size is larger than 40 for the case with perfect compliance and no loss of follow up. \par

\subsection{Sample size Calculation for Two Arms Clinical Trials with Ordinal Categorical Endpoints}\label{sec2.5}

The function \textsf{getSizeOrd()} calculates the required sample size for two arms clinical trial with ordinal categorical end points and needs the following ten input parameters to estimate the plausible sample size:\par 

For a given two arms trial, \textsf{design} is the allocation method (parallel/crossover), \textsf{test} is the hypothesis test(equality/noninferiority/superiority/equivalence), \textsf{alpha} is the level of test significance, \textsf{beta} is the type II error (1-power), \textsf{varcatprob} is the list of two probability vectors (one per treatment arm), \textsf{k} is the ratio of control arm size to the treatment arm size, \textsf{delta} is the delta margin in the test hypothesis, \textsf{theta} is the log odds ratio of outcome in treatment arm versus control arm, \textsf{rho} is a vector of length two with noncompliance rates of control arm and treatment arm, and,  \textsf{r} is the projected proportion of trial uniform loss of follow-up.  The function general code in R is: \par 
\begin{center}
\begin{Verbatim} 
getSizeOrd(design,test,alpha,beta,varcatprob,k,theta,delta,rho,r)
\end{Verbatim}    
\end{center} 
As the last working example, we consider the Example 1 \cite{Whitehead1993} with additional features of noncompliance and expected loss of follow up:\par
\textbf{Example 4.} A pharmaceutical company is exploring the possibility of conducting a two-arm parallel clinical trial to test equality of patient response outcomes three months post-treatment. The survey categorizes responses as "very good," "good," "moderate," and "poor." The principal investigator (PI) is presently determining the required sample size for a trial with 90\% power at a 5\% significance level. The calculation is based on assumed response rates of 0.2, 0.5, 0.2, and 0.1 for "very good," "good," "moderate," and "poor" outcomes in the control arm, respectively. In the treatment arm, the assumed response rates are 0.378, 0.472, 0.106, and 0.044, respectively. The PI is considering equal-sized arms, a log odds ratio of 0.887, and factoring in noncompliance rates of 5\% for the control arm and 7\% for the treatment arm. Additionally, the PI is accounting for an overall trial loss to follow-up of 10\%. The R code for sample size calculations given above trial information is given by:\par 
\begin{Verbatim} 
line #1:>  getSizeOrd(design="parallel",
                      test="equality",
                      alpha=0.05,
                      beta=0.10,
                      varcatprob= list(c(0.2,0.5,0.2,0.1),
                                       c(0.378,0.472,0.106,0.044)), 
                       k=1,
                       theta=0.887,
                       delta=0,
                       rho=c(0.05, 0.07),
                       r=0.1)
\end{Verbatim} 
yielding to following output estimations:
\begin{Verbatim}
line #1:>       n_2 n_1
line #2:> Size   135   135    
\end{Verbatim}
Hence, the PI needs to consider two arms each of size 135. We note that this sample size is larger than 94 for the case with perfect compliance and no loss of follow up. \par

\section{Sample Size Variation by Noncompliance and Loss of Follow-up}\label{sec3}

In all the examples presented above, it is evident that the necessary sample size is consistently larger when dealing with noncompliance and loss of follow-up compared to scenarios involving perfect compliance and no loss of follow-up. A crucial question arises regarding how the estimated sample sizes fluctuate across different values of noncompliance rates and loss of follow-up. Furthermore, whether there is practical application of these considerations at the time of trial  design.  The subsequent subsections delve into a detailed discussion of these aspects. In the upcoming discussion we assume equal arms of noncompliance rates following up the well-known Fibonacci numbers \cite{Debnath2011,Kalman2003} (i.e., 0\%, 1\%, 2\%, 3\%, 5\%, 8\%, 13\%) and expected loss of follow up rates in the range of 0\%-20\% with increments of 5\%. We remind the reader that these numbers are chosen hypothetically while in past conducted clinical trials,  their true empirical rates were found close to of one these hypothetical points. Some empirical examples are found in \cite{Chaudhryetall2020,Besch1995,Czobor2011,Cuzick1997,Boudes1998}.\par

\subsection{Example: Secondary ITT  Estimate vs. Primary ITT Estimate}\label{sec3.1}
A team of investigators at Endologix Inc conducted a two-arm parallel randomized control clinical trial to assess the safety and efficacy of the new proposed AFX Endovascular system  versus comparator proximally-fixed Endovascular Aneurysm Repair (EVAR) system. These medical devices were used in treatment of patients aged 64-80 with infrarenal Abdominal Aortic Aneurysms (AAA) \cite{Kwolektal2023}.  The trial named "Looking at EVAR Outcomes by Primary Analysis of Randomized Data" (LEOPARD) was planned in 80 centers in the USA.  Treatment regimen assignment of AFX device to the comparator device was set in 1:1 ratio. The investigators needed a sample size to achieve a power of 80\% (one-sided significant level: 0.05) to show superiority of AFX versus the comparator. The target treatment response rates for AFX arm and the comparator arm were assumed 86\% and 79\%, respectively. The original sample size estimate was 724 patients. However, the trial PIs decided to augment the sample size to 804 patients justifying it for 10\% drop out rate (as combination of loss of follow-up, withdrawal of informed consent, and death after one month in the trial).\par
\begin{figure}[H]
\includegraphics[clip,width=\columnwidth,height=0.80\textwidth]{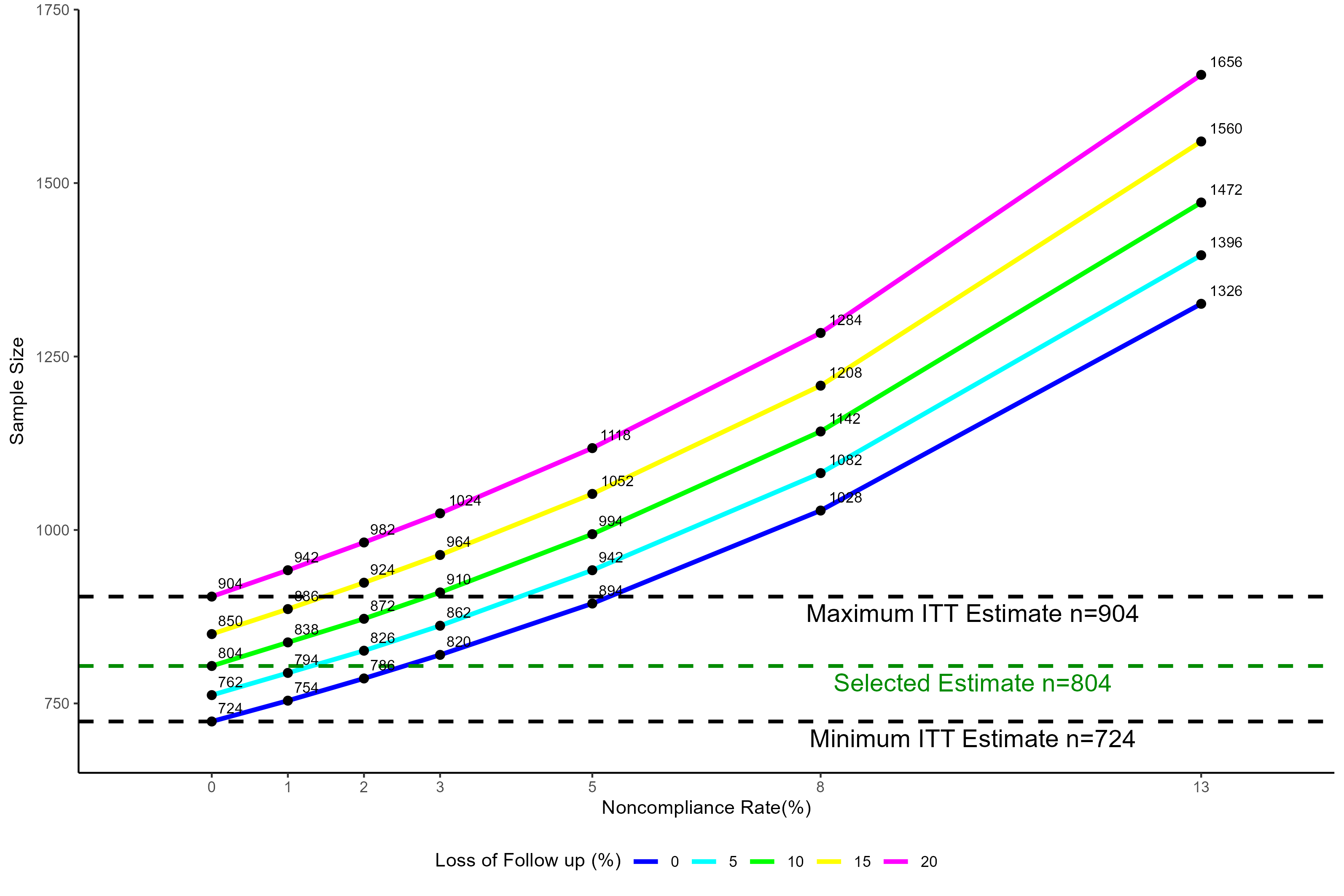}%
\caption{ Estimates of total sample size  for binary end point scenario in  Endologix Inc LEOPARD trial  in terms of equal arms noncompliance rates and loss of follow up\label{fig2}}
\end{figure}

 
Here, we propose an comprehensive Secondary ITT- Primary ITT exploration for this logical decision to increase of sample size given that the trial PIs  do  not know the precise values for noncompliance and loss of follow up. They consider  equal arms non compliance rates   0\%, 1\%, 2\%, 3\%, 5\%, 8\%, 13\%   and expected loss of follow up rates of  0\%, 5\%, 10\%, 15\%, 20\%. Figure \ref{fig2} presents the estimation and variation of required sample sizes. As it is shown, within each level of loss of follow up with higher noncompliance rates study PIs require  higher sample size to achieve the same 80\% of power and target treatment effect. Mathematically, this is due to the fact that by equation (\ref{eq11}) we have $V(p_1^{*},p_2^{*},\delta)=p_2^{*}-p_1^{*}-\delta= (1-\rho_2-\rho_1)\times (p_2-p_1)-\delta=0.07(1-2\rho) \downarrow$ when $\rho \uparrow.$ Thus, by equation (\ref{eq6}) we have $n_2^{*}\uparrow,$ in equation (\ref{eq14}).\par 
Secondly, the study PIs assumed an ideal ITT scenario with required sample size of 804. Had the study shown a true equal arms non compliance rates of 3\%, the study needed a sample size of 910 to maintain its 80\% power. Ignoring  such required increase in sample size and providing the original ITT sample size of 804, would yield to an underpowered  study with the estimated power of 75.5\%.\par 
Finally, a close look at the bottom left side of Figure \ref{fig2} reveals three  candidate estimate sample sizes to the adopted final estimate sample size of 804: (i) 804 (loss of follow up: 10\%; noncompliance: 0\%); (ii) 794 (loss of follow up: 5\%; noncompliance: 1\%); and (iii) 786 (loss of follow up: 0\%; noncompliance: 2\%). \par

\vspace{-6pt}

\begin{remark}\label{uequalrates}
In the above comprehensive Secondary ITT- Primary ITT exploration the study PI assumed the ideal scenario of equal arms noncompliance rates. However, due to various factors there may be unequal arms noncompliance rates. Table \ref{table2} presents the case of three scenario estimations for the expected loss of follow up rate of 10\%. As it is shown, similar to observation above and within each  scenario (fixed row), as the arms associated noncompliance rates increase, the required trial total sample size increase as well. Furthermore, as we move from $\rho_1\leq\rho_2$ to $\rho_1=\rho_2$ and then from $\rho_1=\rho_2$ to $\rho_1\geq\rho_2$ (fixed column), the required trial total sample size follows an increasing-decreasing trend. This trend suggests that the scenario $\rho_1=\rho_2$ yields maximum required sample size in all three scenarios within the fixed column. Consequently, it appears the assumption of equal arms noncompliance rates yields to the conservative (i.e., maximum) sample size calculation. We theoretize this observation in the following conjecture:
\end{remark}

\begin{table}[H]
\caption{Estimates of total sample size for binary end point scenario in Endologix Inc LEOPARD trial in terms of arm related noncompliance rates status.\label{table2}} %title of the table
\footnotesize{
\begin{threeparttable}[b]
\centering % centering table
\begin{tabular}{p{1.25cm}p{1.25cm}p{1.25cm} p{1.25cm}p{1.25cm}p{1.25cm}p{1.25cm}p{1.25cm}} % creating four columns
\toprule % inserts single-line
\textbf{Scenario} & &  & & $n(\rho_1,\rho_2)$ & & & \\ % Entering row contents
\midrule % inserts single-line
$\rho_1\leq\rho_2$ & $n(0,0)=$ &$n(0,1)=$ &$n(1,2)=$ &$n(2,3)=$ &$n(3,5)=$ &$n(5,8)=$ &$n(8,13)=$\\
& 804 &822 &856 &892 &954  &1068 &1302 \\ 
$\rho_1=\rho_2$    & $n(0,0)=$ &$n(1,1)=$ &$n(2,2)=$ &$n(3,3)=$ &$n(5,5)=$ &$n(8,8)=$ &$n(13,13)=$\\
& 804 &838 &872 &910 &994&1142& 1472  \\ 
$\rho_1\geq\rho_2$ & $n(0,0)=$ &$n(1,0)=$ &$n(2,1)=$ &$n(3,2)=$ &$n(5,3)=$ &$n(8,5)=$ &$n(13,8)=$\\
& 804 &818 &854 &890 &948  &1058 &1282 \\ 
\bottomrule % inserts single-line
\end{tabular}
\end{threeparttable}
\begin{tablenotes}
\footnotesize
\item[a] Notes: $n$ refers to total trial sample size. Numerical values for $\rho_1,\rho_2$ are in percentage scale. 
\end{tablenotes}
}
\end{table}

\begin{conjec}\label{conj1}
Let $n^{*}=(1+k)n_{2}^{*}$ be the be the total two arm clinical trial sample size with $n_{2}^{*}$ given by one of the equations (\ref{eq13})-(\ref{eq16}). Then, for fixed values of trial arm treatment parameters and expected loss of follow up rate, $n^{*}$ attains its maximum if and only if for some bivariate closed form real-valued function $c(.,.)$ the trial PI assumes   $c(\rho_1,\rho_2)=0.$
\end{conjec}

\subsection{Statistical Software Role in Estimations: Commercial vs. Noncommercial}\label{sec3.2}
\begin{figure}[H]
\includegraphics[clip,width=\columnwidth,height=0.8\textwidth]{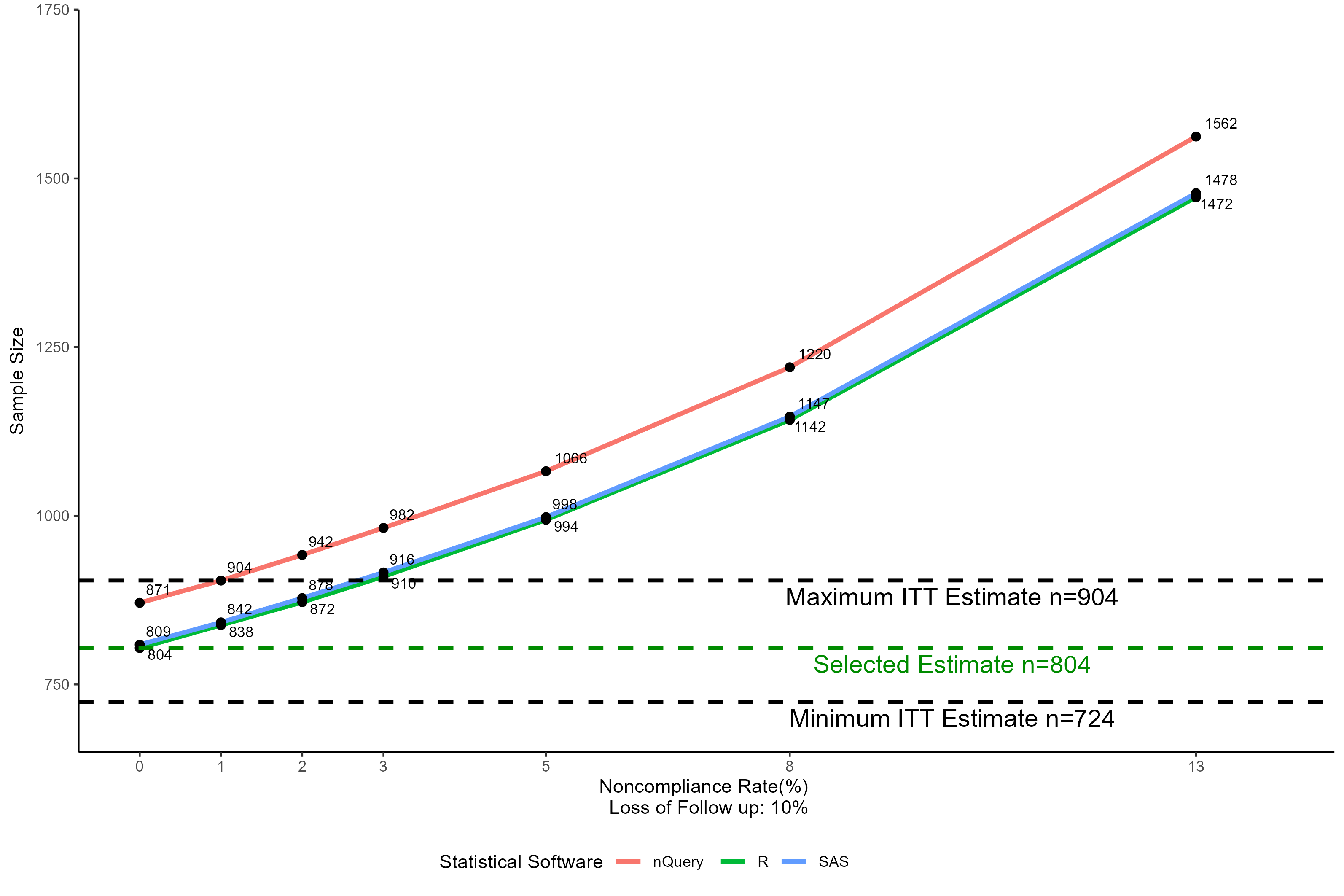}%
\caption{ Estimates of total sample size for binary end point scenario in Endologix Inc LEOPARD trial  in terms of noncompliance rates and statistical software\label{fig3}}
\end{figure}

We discuss the role of statistical software in estimations of total sample size. Referring to previous section and the  best candidate i.e., 804 (loss of follow up: 10\%; noncompliance: 0\%) we compare its  associated estimation over span of positive noncompliance rates and across  three statistical softwares: R (package: GenTwoArmsTrialSize 0.0.5: Wald test with unpooled variance); nQuery (version 7.0; Fisher's Exact test) \cite{nQuery2007}; and SAS (version 9.4: Farrington-Manning Score test) \cite{SAS2023}. As it is shown in Figure \ref{fig3}, for the fixed loss of follow up of 10\%, R and SAS statistical softwares present similar estimations for the total trial sample size for the noncompliance rates of 0-13\%.\par      

\vspace{-6pt}

\section{Discussion}\label{sec4}

\subsection{Summary $and$ Contributions}\label{sec4.1}

We present GenTwoArmsTrialSize, an R package that stands as the inaugural statistical software tool for determining sample sizes in two-arm clinical trials, incorporating considerations for noncompliance and loss of follow-up. This package offers the ability to calculate sample size considering (1) four types of endpoints, namely continuous, binary, time-to-event, and ordinal categorical, (2) both crossover and parallel designs, and (3) hypothesis tests for equality, noninferiority, superiority, or equivalence. Moreover, the uniformity of the estimating function's application across all eight scenarios for each endpoint enhances the user-friendliness of the estimation process for pharmaceutical statisticians. We anticipate that this package will assist researchers in comprehensively, precisely, and efficiently estimating the required sample size. Additionally, it provides valuable insights for pharmaceutical statisticians involved in trial design by illustrating the impact of noncompliance and loss of follow-up on the variability of the required sample size. \par

\subsection{Limitations $and$ Future Work}\label{sec4.2}
The limitations of this work are clear and each on its own creates opportunities for extra research in sample size calculations as follows:\par 
First, as in previous presented R-packages, Scenarios 9--12 (small sample size with continuity correction), 21--24 and 29--32 in the Table \ref{table2} are missing in the current package and need further development in subsequent versions.\par  
Second, we did not cover the bioequivalence studies design in the current version \cite{Labesetall2022}. Given wide spectrum of bioequivalence trial designs, they demand their own special package addressing noncompliance and loss of follow-up in the sample size estimations. However,  the current package still is useful as follows: Consider the following bioequivalence hypothesis testing:\par 
\begin{eqnarray}
Bioequivalence (1) &:&  H_0:  \theta_2-\theta_1 \leq \delta_1\  or\  \theta_2-\theta_1\geq \delta_2\ \ \ vs.\ \ \     H_1:  \delta_1<\theta_2-\theta_1<\delta_2,\label{eq4bioeq1}\\
Bioequivalence (2) &:&  H_0: \frac{\theta_2}{\theta_1}\leq \delta_1\  or\  \frac{\theta_2}{\theta_1}\geq \delta_2\ \ \ vs.\ \ \     H_1:  \delta_1<\frac{\theta_2}{\theta_1}<\delta_2.\label{eq4bioeq2}
\end{eqnarray}

Then, with some mathematical calculation we may write equations (\ref{eq4bioeq1}) and (\ref{eq4bioeq2}) in the form of Equivalence test equation (\ref{eq4}) via:
\begin{eqnarray}
Bioequivalence (1) &:& H_0: |\theta_2^{*}-\theta_1^{*}|\geq +\delta^{*}\ \ \ vs.\ \ \     H_1: |\theta_2^{*}-\theta_1^{*}|<+\delta^{*}, \label{eq4bioeq1b}\\
&:& \theta_2^{*}=\theta_2-\frac{\delta_2}{2}, 
\theta_1^{*}=\theta_1+\frac{\delta_1}{2}, 
\delta^{*}=\frac{\delta_2-\delta_1}{2} ,\nonumber\\
Bioequivalence (2) &:& H_0: |\theta_2^{**}-\theta_1^{**}|\geq +\delta^{**}\ \ \ vs.\ \ \     H_1: |\theta_2^{**}-\theta_1^{**}|<+\delta^{**},\label{eq4bioeq2b}\\
&:& \theta_2^{**}=\log(\theta_2\delta_2^{-\frac{1}{2}}), \theta_1^{**}=\log(\theta_1\delta_1^{\frac{1}{2}}),  \delta^{**}=\log(\frac{\delta_2}{\delta_1})^{\frac{1}{2}}.\nonumber 
\end{eqnarray}
Accordingly,  all after-mentioned estimations in the current version of the package maybe utilized. \par 
Third, a natural extension of the two arms trial design is the multiple arms trial design. One key consideration in such extension is the multiple testing correction in the process of the estimation yet there are various perspectives on it \cite{Wasonetal2014,Wasonetal2016}.\par 
Fourth, there are other clinical trial designs that additional features of noncompliance and loss of follow-up in their sample size estimation methods need attentions including multicenter trials \cite{Hayes1999}, cluster trials \cite{ Heoatal2016}, longitudinal trials \cite{Iddi2022} and their combinations \cite{Ahnetal2014}. \par 
Fifth, in this work we considered adjustment for sample size calculation for loss of follow up and noncompliance in two sequential stages. However, some cases such as in equations (\ref{eq13}-\ref{eq16}), each of these adjustments have been addressed in one stage \cite{Julious2023}. Comparison of these two methods estimations remains an open problem.\par 
Sixth, this work did not consider standardized effect size. In the real clinical trial design setting, there are always some random variation in the assumed treatment estimation for each trial arm, which ignoring in the calculation of sample size yields imprecise estimations. This feature will be considered in the next update of this R package.\par
Seventh, the current package can be extended to estimate the power of the two arms clinical trial in terms of above after mentioned key trial features and sample size as well. In details, let $f(.)$ be one of key functions $getSizeMean(.), getSizeProp(.),getSizeTTE(.),$ or $getSizeOrd(.).$ Then, for the given power $\pi$ and vector of other features $\Vec{x}$ we have $n=f(1-\pi,\Vec{x}).$ Defining a function $h(.)$ via $h(\pi,\Vec{x},n)=f(1-\pi,\Vec{x})-n,$ we have $\hat{\pi}=g(n,\Vec{x})=\{\pi\in[0,1]|h(\pi,\Vec{x},n)=0\}.$ The solution for the last equation is derived using the function ``uniroot" in the R stats package \cite{RCoreTeam2023} (See Appendix \ref{secB}).\par 
Eighth, this R package only accounts for fixed designs in its calculations. However, in conducting clinical trials, many Institutional Review Boards (IRBs) require some form of early stopping for futility, which limits the package's applicability in these scenarios. Therefore, practicing pharmaceutical statisticians desire its generalization to adaptive designs.\par      
Finally, the proof or disproof for the mathematical conjecture \ref{conj1} remains an open problem.\par 

\subsection{Conclusions}\label{sec4.3}

The GenTwoArmsTrialSize R package, available at no cost, empowers users to determine the necessary sample size for two-arm trials across a comprehensive range of scenarios. These estimations enable statisticians to account for various factors, such as noncompliance and loss of follow-up, enhancing precision in their calculations. This heightened precision, in turn, allows researchers to conduct sample size calculations that are not only more economically sound but also more ethically and scientifically grounded.\par

%%%%%%%%%%%%%%%%%%%%%%%%%%%%%%%%%%%%%%%%%%%%%%%%%%
\subsection*{ORCID}
\begin{table}[H]
\begin{tabular}{ll}
Mohsen Soltanifar: & 0000-0002-5989-0082   \\  
Chel Hee Lee: & 0000-0002-8244-374X  \\
Amin Shirazi: & 0009-0009-3091-9492   \\ 
Martha Behnke: & 0009-0002-1139-6963  \\  
Ilfra Raymond-Loher: & 0009-0000-3453-3276 \\ 
Getachew A. Dagne: & 0000-0002-7052-2221  
\end{tabular}
\end{table}

\subsection*{Supplementary}
The following supporting information can be downloaded at ``GenTwoArmsTrialSize'' R package CRAN documentation: \url{https://CRAN.R-project.org/package=GenTwoArmsTrialSize}. 

\subsection*{Author Contributions}
Conceptualization, M.S.; methodology, M.S.; software, M.S. and C.H.L.; validation, M.S.,  C.H.L., A.S., M.B, and I.L ; formal analysis, M.S.; investigation, M.S.,  C.H.L., A.S., M.B, I.L and G.A.D; resources, M.S. and C.H.L.; data curation, M.S.; writing---original draft preparation, M.S., C.H.L., A.S., M.B, I.L, G.A.D; writing---review and editing, M.S., C.H.L., A.S., M.B, I.L, G.A.D; visualization, M.S. and C.H.L.; supervision, M.S. and C.H.L.; project administration, M.S. and C.H.L.; funding acquisition, M.S. All authors have read and agreed to the published version of the manuscript.

\subsection*{Funding}
This paper  has no external funding.

\subsection*{Institutional Review}
Not applicable.

\subsection*{Informed Consent}
Note Applicable.

\subsection*{Data Availability}
Not applicable.

\subsection*{Acknowledgments}
Not Applicable.

\subsection*{Conflicts of Interest}
The authors declare no conflict of interest. 

\subsection*{Abbreviations}
The following abbreviations are used in this manuscript: \\
AAA: Abdominal Aortic Aneurysms; 
allo: allogeneic; auto: autologous; CACE: Complier Average Causal Effect; CDF: Cumulative Distribution Function; CHD: Chronic Heart Disease; CRAN: Comprehensive R Archive Network; EVAR: Endovascular Aneurysm Repair;  ICH: International Council for Harmonisation of Technical Requirements for Pharmaceuticals for Human Use; IRB: Institutional Review Board; ITT:  Intention To Treatment; LDL: Low Density Lipoprotein; LEOPARD: Looking at EVAR Outcomes by Primary Analysis of Randomized Data; LoF: Loss of Follow-up;  OR: Odds Ratio; Noncom: Noncompliance; PI: Principal Investigator; RCT: Randomized Clinical Trial; SD: Standard Deviation; TTE: (a) Time to Event; (b) Target Treatment Effect.  

%%%%%%%%%%%%%%%%%%%%%%%%%%%%%%%%%%%%%%%%%%%%%%%%%%

\section*{Appendix}
\subsection*{Mathematical Proofs:}\label{secA}
We outline the alternative proofs for equations (\ref{eq9})-(\ref{eq11}) outlined before in \cite{Jo2002}. Let $Z_i\sim Bernoulli(\rho_i)$ be the noncompliance random variable in the ith trial arm $(i=1,2).$ Then, by the law of iterated expectation we have:
\begin{eqnarray}
 \theta_{i}^{*}&=& E(\hat{\theta}_i)\nonumber\\
               &=& E(E(\hat{\theta}_i|Z_i))\nonumber\\
               &=& P(Z_i=0) E(\hat{\theta}_i|Z_i=0) +P(Z_i=1) E(\hat{\theta}_i|Z_i=1) \nonumber\\
               &=& (1-\rho_i) \theta_i + \rho_i \theta_{(3-i)}:\ \ \ \  i=1,2. \nonumber
\end{eqnarray}
This completes the proof for the equations (\ref{eq9})-(\ref{eq10}). Next, a subtraction of equation (\ref{eq10}) from equation (\ref{eq9}) yields:
\begin{eqnarray}
 \theta_{2}^{*}- \theta_{1}^{*}&=& (\rho_2\theta_1+(1-\rho_2)\theta_2)-((1-\rho_1)\theta_1+\rho_1\theta_2) \nonumber\\    
 &=& (\rho_2-1+\rho_1)\theta_1+(1-\rho_2-\rho_1)\theta_2\nonumber\\
 &=& -(1-\rho_1-\rho_2)\theta_1+(1-\rho_1-\rho_2)\theta_2 \nonumber\\
 &=& (1-\rho_1-\rho_2) (\theta_2-\theta_1).\nonumber 
\end{eqnarray}
This completes the proof for the equation (\ref{eq11}).

\subsection*{Power Calculation Example:}\label{secB}
We consider Example 1 in section \ref{sec2.2}. The study PI considers equal arms sample size of 113. The  power of the trial with same conditions as in there will be:

\begin{Verbatim} 
line #1:>  library(GenTwoArmsTrialSize)
line #2:>  TrtAllocation_1 <-"parallel"
           HypothesisTest_1<-"equivalence"
           alpha_1 <- 0.05
           sigma_1 <- 0.10
           k_1 <- 1
           delta_1 <- 0.05
           TTE_1 <- 0.01
           rho_1 <- c(0.05,0.07)
           r_1 <- 0.1
           n_1<-113
line #3:>  h <- function(beta, n){
                 n0 <-getSizeMean(design=TrtAllocation_1,
                             test=HypothesisTest_1,
                              alpha=alpha_1, beta=beta, sigma=sigma_1, k=k_1,
                              delta=delta_1, TTE=TTE_1,
                              rho=rho_1, r=r_1);
                 n0 <- as.numeric(n0[1]);
                 v <- n0 - n;
                 return(v)}
line #4:> power=round(1-uniroot(h,interval=c(0.01, 0.99), 
                                  n=as.numeric(n_1[1]))$root,5)
line #5:> power 
line #6:> [1] 0.80138 
\end{Verbatim}

%%%%%%%%%%%%%%%%%%%%%%%%%%%%%%%%%%%%%%%%%%%%%%%%%%%

\end{document}